\documentclass[%
 reprint,
 amsmath,amssymb,
 aps,
]{revtex4-2}

\usepackage{graphicx}
\usepackage{dcolumn}
\usepackage{bm}
\usepackage{xspace}
\usepackage{dsfont}

\usepackage{braket}
\usepackage{amsmath}

\begin{document}

\preprint{APS/123-QED}

\title{Exact results on Quantum search algorithm}

\author{Saptarshi Roy Chowdhury}
\email{saptarshi@barc.gov.in}
\author{Swarupananda Pradhan}
 
\affiliation{%
 Beam Technology Development Group, Bhabha Atomic Research Centre, Mumbai-400085, India
}

\begin{abstract}
We obtain \emph{exact} analytic expressions for the success probability and the dynamics of the off-diagonal terms in the density matrix after arbitrary number of iterations of the generalized Grover operator with two generic phase angles $(\alpha, \beta)$. Using this we find for the phase matching condition $\alpha=-\beta=0.268\pi$ with \emph{three} iterations, we can achieve success probability $\geq0.8$ only with a knowledge about the lower bound $\lambda=0.14$ where $\lambda$ is the ratio of marked to total number states in the database. Finally we quantify success probability of the algorithm against initial state preparation errors using a simple model.

\end{abstract}

\keywords{Suggested keywords}
                              
\maketitle

\section{\label{sec:level1}Introduction
\protect\\ 
 \lowercase{} }
There has been a constant effort since few decades in order to harness the power of Quantum mechanics in various fields of science and technology. The subject of Quantum computation is one such example which is believed to hold exceptional promise in the sense of solving some of the crucial problems with significantly better computational advantage than their classical counterpart. Quantum algorithms such as Deutsch–Jozsa algorithm \cite{deutsch1985quantum}, Shor's factorization algorithm \cite{365700}, Grover's quantum search algorithm \cite{PhysRevLett.79.325,10.1145/237814.237866} are some of the notable examples in this regard. Grover's quantum search gives square root speed up over the best available classical algorithm when searching for marked items from an unstructured database. 

After this seminal finding, several efforts have been made to extend and improve the quantum search algorithm. Bennett \textit{et al.} \cite{bennett1997strengths} proved square root speed up to be computationally optimal, also see the work by Zalka \cite{PhysRevA.60.2746}. It was shown to be a larger class of quantum amplitude amplification problem by Brassard and Hoyer \cite{Brassard:1997gj}. Long 
 \cite{PhysRevA.64.022307} generalized the algorithm for generic phase angles and gave exact phase expressions for $100\%$ success.  Boyer \textit{et al.} \cite{1998ForPh..46..493B} showed how to handle multiple marked states in the database. Biham \textit{et al.} \cite{PhysRevA.63.012310,PhysRevA.60.2742,PhysRevA.66.062301} found success probability for arbitrary iteration with the original phase matching for  a general pure and mixed initial state. For some of the further results, see \cite{PhysRevA.65.052322,PhysRevA.65.034305,article}. Li and Li \cite{LI200742} reported for $\lambda\geq1/3$, where $\lambda$ is the ratio of the marked to total number of states in the database, success probability $P(\lambda)\geq25/27$ can be achieved with a single iteration. Multi-phase matching condition was proposed and numerical results were given in order to improve the success probability for wide range of \(\lambda\) \cite{toyama2009matched,PhysRevA.77.042324}. Dependence of  coherence was studied in \cite{PhysRevA.95.032307} using several typical measures of quantum coherence and quantum correlations for pure states using previous results \cite{PhysRevA.63.012310,PhysRevA.60.2742,PhysRevA.66.062301}. For the application of quantum search ideas to fermionic systems, see the recent paper by Roget \textit{et al.} \cite{PhysRevLett.124.180501}. Grover iteration can be thought of as a rotation in the two-dimensional Hilbert space spanned by marked and unmarked states \cite{preskill1998lecture}. Each rotation slowly transforms the initial state which is a uniform superposition towards the marked states. Unitarity of the Grover operator implies there will always be so called `overcooking'(`under cooking') of the  prepared state if we iterate more than the required number as the eigen values are inherently periodic which often poses difficulty when number of marked items is unknown. To get around this problem, the idea of fixed point quantum search was proposed by Grover \cite{PhysRevLett.95.150501} where the success probability always gets an improvement with each iteration as the iteration operator follows a recursion relation implies each iteration is not the identical unitary which would've prevented the algorithm to have a fixed point earlier. Fixed point quantum search is useful when \(\lambda\) is unknown at the price of decreased efficiency of the algorithm \cite{PhysRevA.72.032326}. Recently Chuang \textit{et al.} \cite{PhysRevLett.113.210501} claimed a fixed point search involving a different phase matching condition with optimal number of user controlled oracle queries using functions typically used as frequency filters in electronics. Some of the experimental realizations of Grover algorithm have also been reported, see for example \cite{PhysRevLett.80.3408,PhysRevA.66.042310,PhysRevA.72.032326,PhysRevA.72.050306,PhysRevApplied.7.054025}.

 In this work, we generalize Grover algorithm in a density matrix set up. We find \emph{exact} success probability and the dynamics of the off-diagonal term in the density matrix for arbitrary iteration as a function of number of iterations, two generic phase angles$(\alpha,\beta)$ and parameter $\xi$ introduced in the off diagonal terms of the density matrix in order to capture the coherence present in the initial quantum register. We use the success probability expression to show for the phase matching condition $\alpha=-\beta$ with various phases and iterations, we can achieve  success probability profile which is always greater than a fixed threshold value only with a knowledge about lower bound on $\lambda$. In particular, we show with \emph{three} iterations and $\alpha=-\beta=0.268\pi$, we can achieve $P(\lambda)\geq0.8$ with lower bound to be $\lambda=0.14$. We tabulate the values of phase angle $\alpha$ and the lower bound on $\lambda$ for which we consistently get $P(\lambda)\geq0.8$ and $\geq0.9$ with successive iterations. This fixes the problem of so called `overcooking'(`under cooking') of the state for unknown number of marked states as precise knowledge of $\lambda$ is not required except for a lower bound and also having to do with an user controlled oracle query suggested recently \cite{PhysRevResearch.4.L022013}. Finally, We quantify how success probability of the algorithm gets affected with the initial state preparation errors for various phase matching conditions given for different limits of $\xi$ in this simple model. 

The paper is organized as follows. In Section II A, we give a brief overview of the Grover algorithm. Form of the most general operators with two generic phase angles originally given by Li and Li \cite{LI200742} is given with their results in Section II B. In Section III we present our \emph{exact} analytic results for the success probability and the dynamics of the off-diagonal term in the density matrix for arbitrary number of iteration as a function of the parameters of the algorithm. Section IV contains further results from the success probability expressions particularly extension of the \cite{LI200742} phase matching condition to a lesser lower bound on $\lambda$ and quantification of the success probability against initial state preparation errors. We end the paper with a discussion on the result in Section V. 
\section{Preliminaries}
We review Grover algorithm in its original form. We point out the two dimensional subspace spanned by the marked and unmarked state which simplifies the discussion of the algorithm. Proper generalization of the algorithm in terms of two generic phase angles and the result of \cite{LI200742} is discussed after that.   

\subsection{Overview of Grover Algorithm}

Suppose we have an unstructured database of size \(N=2^n\) and want to search for M marked items from the database. Classically, query complexity of the problem scales as $O(N/2M)$ with the size of the database. Quantumly, as Grover showed in his remarkable paper \cite{PhysRevLett.79.325,10.1145/237814.237866}, we can achieve quadratic speed up over the classical case. Grover algorithm consists of initialization of the $n$-qubit state $\ket{0}^{{\otimes}\,n}$ to equal superposition state using $n$-qubit Hadamard gate as $ \ket{\psi}=H^{{\otimes}\,n}\ket{0}^{{\otimes}\,n}=\frac{1}{\sqrt{N}}\sum_{x=1}^{2^n}\ket{x}$,  Apply the following two operators: i) Oracle query $O=(1-2\sum_{x=0}^{M}\ket{x}\bra{x})$, ii)  Diffuser operator $D=(2\ket{\psi}\bra{\psi}-1)$ iteratively \(k_{opt}=(\pi/2\theta-1/2\)) times with \(\theta\) given by, \(\theta=2\arcsin\sqrt{\lambda}\). Defining Grover operator $G$ as $DO$, effect of $G$ on \(\ket{\psi}\) is essentially captured by the rotation matrix written in the basis \(\ket{R},\ket{T}\) as,
\[ G=
         \begin{pmatrix}
             \cos\theta & -\sin\theta \\
             \sin\theta & \cos\theta
         \end{pmatrix}
\]
with the bases \(\ket{R}\) and \(\ket{T}\) given by
\[\ket{T}=\frac{1}{\sqrt{M}}\sum_{x=1}^{M}\ket{x}, \;\;
\ket{R}=\frac{1}{\sqrt{N-M}}\sum_{x=M+1}^{N}\ket{x}
\]
respectively the uniform superposition of marked and unmarked states. In terms of these bases, initial state can be written as,
\[ \ket{\psi}=\sqrt{1-\lambda}\ket{R}+\sqrt{\lambda}\ket{T}
\]
Application of Grover operator has the effect of rotating $\ket{\psi}$ towards \(\ket{T}\) through an angle \(\theta\) in each iteration taking \(k_{opt}\) iteration in total before measurement can be done in the computational basis completing the quantum search process. Exact knowledge of $\lambda$ is required for the algorithm to succeed as slight mismatch of the number of iterations from $k_{opt}$ could lead to significant decrease in performance due to inherent periodicity of the success probability. 
\subsection{Generalization of the algorithm with generic phases}
Generalization of the Grover operator has been made with two generic phases written in the basis of \(\ket{R},\ket{T}\) as \cite{LI200742},
\begin{align}
U(\alpha)=I-(1-e^{i\alpha})\ket{T}\bra{T} \\
V(\beta)=Ie^{i\beta}+(1-e^{i\beta)}\ket{\psi}\bra{\psi}
\end{align}
where $U$ selectively shifts the phases of the marked states by an angle \(\alpha\) and $V$ shifts the phase by angle \(\beta\) around the fixed state \(\ket{\psi}\) each time. This reduces to the original Grover algorithm for the choice $\alpha=\beta=\pi$.

Li and Li gave their new phase matching condition as $\alpha=-\beta=\frac{\pi}{2}$. This gives the result that with single iteration, we get success probability $P(\lambda)\geq\frac{25}{27}$ for $1/3\leq\lambda\leq1$. A geometric picture on this particular phase matching in terms of three independent vectors can be found here \cite{LI200742}.  

\section{Exact expressions with Generalized phase angles}
We begin with the matrix representation of the operator $G(\alpha,\beta)=U(\alpha)V(\beta)$ :
\[
\begin{pmatrix}
    (1-e^{i\beta})(1-\lambda)& e^{i\alpha}(1-e^{i\beta})\sqrt{\lambda(1-\lambda)}\\
     (1-e^{i\beta})\sqrt{\lambda(1-\lambda)}&e^{i\alpha}(1-e^{i\beta})\lambda+e^{i(\alpha+\beta)}
\end{pmatrix}
\]
Let the initial density matrix be
\begin{multline}
\rho_{initial}=(1-\lambda)\ket{R}\bra{R}+\lambda\ket{T}\bra{T}\\
+\xi\sqrt{\lambda(1-\lambda)}(\ket{R}\bra{T}+\ket{T}\bra{R})
\end{multline}
where $\xi$ is a generic parameter introduced to quantify the coherence present in the initial density matrix. 
Density matrix after $m$ iterations is given by,
\[
\rho_{m}=G^{m}\rho_{initial}(G^{\dag})^m
\]
The success probability can be given as the $\bra{T}\rho_{m}\ket{T}$ matrix element of the final density matrix. We raise the $G(\alpha,\beta)$ to the $m^{th}$ power using the standard method of expanding the matrix in the basis of $I$ and the Pauli matrices \cite{PhysRevResearch.4.L022013}. We obtain the success probability as a function of $m$, $\xi$ and generic angles $\alpha$ and $\beta$ as,
\begin{multline}
P(\lambda,\xi,\alpha,\beta,m)= \lambda+\sin^2({m\phi})[(1-n_3^2)(1-2\lambda)]\\
-2\xi\sqrt{\lambda(1-\lambda)}\sin^2({m\phi})[n_1n_3+n_2\cot({m\phi})]
\end{multline}
with,
\begin{align}
\cos(\phi)&=\cos\left(\frac{\alpha+\beta}{2}\right)+2\lambda\sin(\beta/2)\sin(\alpha/2)\\
n_1&=-\frac{\sqrt{\lambda(1-\lambda)}}{\sin(\phi)}2\cos(\alpha/2)\sin(\beta/2)\\
n_2&=\frac{\sqrt{\lambda(1-\lambda)}}{\sin(\phi)}(2\sin(\alpha/2)\sin(\beta/2))\\
n_3&=\frac{1}{\sin(\phi)}\left[-\sin\left(\frac{\alpha+\beta}{2}\right)+2\lambda\sin(\beta/2)\cos(\alpha/2)\right]
\end{align}
We find Li and Li's result to be a special case of $P(\lambda,\xi,\alpha,\beta,m)$ in the limit $\alpha=-\beta=\pi/2$ and $m=1$, i.e. $P(\lambda,1,\pi/2,-\pi/2,1)= 4\lambda^3-8\lambda^2+5\lambda$.

We also find the expression for the dynamics of the off-diagonal term in the density matrix with successive iterations in terms of the parameters in the problem. We denote, $C(\lambda,\xi,\alpha,\beta,m)=\frac{\bra{T}\rho_{m}\ket{R}}{\bra{T}\rho_{initial}\ket{R}}$ and find the real and imaginary part of normalized $C(\alpha,\beta,m,\xi,\lambda)$ as,
\begin{multline}
Re\,C(\lambda,\xi,\alpha,\beta,m)=1+\frac{1}{\xi\sqrt{\lambda(1-\lambda)}}[   \sin^2({m\phi})\\
[n_{1}n_{3}(1-2\lambda)-2\xi\sqrt{\lambda(1-\lambda)}(1-n_1^{2})]\\
+\sin({m\phi})\cos{(m\phi)}n_2(2\lambda-1) ]
\end{multline}
\begin{multline}
Im\,C(\lambda,\xi,\alpha,\beta,m)=\frac{1}{\xi\sqrt{\lambda(1-\lambda)}}[ \sin^2({m\phi})\\
[n_2n_3(2\lambda-1)-2n_1n_2\xi\sqrt{\lambda(1-\lambda)} ]\\
+\sin{(m\phi)}\cos{(m\phi)}[n_1(2\lambda-1)
+2n_3\xi\sqrt{\lambda(1-\lambda)}]]
\end{multline}
 
\section{Results}
From the given expression of $P(\lambda,\xi,\alpha,\beta,m)$, we get the success probability for $\alpha=-\beta$ phase matching condition for arbitrary iteration as,
\begin{multline*}
P(\lambda,\xi=1,\alpha=-\beta,m)=\lambda\\+\sin^2(m\phi)\left[\left(1-\lambda^2\frac{\sin^2(\alpha)}{\sin^2(\phi)}\right)(1-2\lambda)\right]\\
+2\xi\lambda(1-\lambda)\frac{\sin^2(m\phi)}{\sin(\phi)}\left[\frac{\lambda\sin^2(\phi)}{\sin(\phi)}+2\sin^2(\alpha/2)\cot(m\phi)\right]
\end{multline*}
We find from the above expression, for m=3, we get success probability to be $\geq0.8$ in the range $0.14\leq\lambda\leq1$ for the above phase matching condition with $\alpha=0.268\pi$. It also gives exact success for $\lambda=0.2965$. For success probability profile to be $\geq0.9$, we get the lower bound on $\lambda=0.229$ with $\alpha=0.234\pi$. Thus only knowledge about the lower bound of $\lambda$ is required in this protocol. Investigation about the dynamics of the real and imaginary part of the off-diagonal term in the density matrix as a function of iterations for various phase matching conditions and its connection with the efficacy of the algorithm will be done in future elsewhere.  

We also find the performance of the algorithm against modest noise captured by the parameter $\xi$. With $\xi$=0, we have $\rho_{initial}=(1-\lambda)\ket{R}\bra{R}+\lambda\ket{T}\bra{T}$. Success probability of the generalized Grove algorithm in this limit comes out to be
\begin{multline*}
P(\lambda,\xi=0,\alpha,\beta)= \lambda+\sin^2(m\phi)[1-n_3^2-2\lambda(1-n_3^2)]
\end{multline*}
with the same $n_3$ and $\phi$ as earlier. We find decrease in the success of the algorithm for most values particularly for $\lambda\leq0.5$ as is generally expected. For example, with this new phase matching, for the value $\lambda=0.2$, success probability becomes almost $60\%$ of its value with $\xi$=1.
\section{Discussions}
Grover algorithm gives substantially improved computational advantage over the available classical algorithms while searching for marked items in an unstructured database which can essentially be used generically for speeding up all sorts of existing searching problems. In this paper, we address the issue of quantum search for unknown number of marked states without using user based oracle queries. We give \emph{exact} closed form expression for success probability and the off-diagonal terms in the density matrix as a function of the various parameters of the algorithm. We look for phase matching conditions which give better success probability profile over wide ranges of $\lambda$ with minimum iteration. We iterate with the phase matching condition $\alpha=-\beta=0.268\pi$ thrice and get a success probability profile consistently over $0.8$ for lower-bound $\lambda\geq0.14$. For $\lambda\geq0.229$, we get $P(\lambda)$ to be more than $0.9$ throughout the entire range of $\lambda$. We believe these results would improve the quantum search algorithm when applied to an unknown database. Some of the interesting future directions would be to explore other phase matching conditions to get a better success probability profile with lesser lower bounds on $\lambda$, better handling of the mixed states in various recently proposed quantum search protocols and understanding the role of off-diagonal terms in the density matrix on the performance of the algorithm. 
\begin{acknowledgments}

We thank Dr. Sudhir Ranjan Jain for his comments on the manuscript. We thank Dr. Archana Sharma and Dr. A.K.Mohanty for supporting the basic research on Quantum information and Computing.

\end{acknowledgments}
\nocite{*}

\bibliography{Grover}

\end{document}